\documentclass[aps,twocolumn,preprintnumbers,amsart]{revtex4}
\usepackage{amssymb}
\usepackage{color}
\usepackage{graphicx}
\usepackage{appendix}
\usepackage{natbib}
\usepackage{float}
\usepackage{hyperref}
\usepackage{amsmath}
\usepackage[tight,FIGBOTCAP]{subfigure}
\usepackage[matrix,frame,arrow]{xypic}

\makeatother

\begin{document}

\title{One way quantum repeaters with quantum Reed-Solomon codes}

\author{Sreraman Muralidharan$^{1,2}$}

\author{Chang-Ling Zou$^{3}$}

\author{Linshu Li$^{3,4}$}

\author{Liang Jiang$^{3,4}$}

\thanks{liang.jiang@yale.edu}

\affiliation{$^{1}$Department of Electrical Engineering, Yale University, New
Haven, CT 06511 USA}

\affiliation{$^{2}$US Army research laboratory, Adelphi MD 20783 USA}

\affiliation{$^{3}$Department of Applied Physics, Yale University, New Haven,
CT 06511 USA}

\affiliation{$^{4}$ Yale Quantum Institute, Yale University, New Haven, CT 06520 USA}

\date{\today}

\begin{abstract}
We show that quantum Reed-Solomon codes constructed from classical
Reed-Solomon codes can approach the capacity on the
quantum erasure channel of $d$-level systems for large dimension $d$. We study the performance
of one-way quantum repeaters with these codes and obtain a significant improvement in key generation rate compared to previously investigated encoding schemes with quantum parity codes and quantum polynomial codes. We also compare the three generation of quantum
repeaters using quantum Reed-Solomon codes and
identify parameter regimes where each generation performs the best.
\end{abstract}
\maketitle

\section{Introduction}
The possibility of transmitting information encoded into quantum states offers unconditional in principle security \cite{LoHoi-KwongandChau1999,Scarani2009,Gisin2007,Shor2000} and can potentially lead to a secure quantum internet \cite{Kimble2008}. There are currently two approaches taken to the transmission of quantum states using single photons \cite{Simon2017}. One uses satellite to 
link remote parties \cite{Ren2017,Yin2017a}, while the other uses fiber based quantum repeaters \cite{Dur1999}. Fiber based quantum repeaters have the potential to offer higher bandwidth, larger key generation rates and is tolerant to inclement weather conditions compared to satellite based quantum communication. 
Long distance classical communication through optical fibers is made possible through establishing intermediate repeater stations, where the optical signal is amplified and retransmitted
to the neighboring station to compensate fiber attenuation. However, quantum communication relies
on the transfer of quantum states - which unlike classical states,
cannot be perfectly amplified or duplicated \cite{Wootters82}. Quantum repeaters
\cite{Briegel1998} (QRs) provide the only known approach for long
distance quantum communication through optical fibers \cite{Takeoka2014,Pirandola2017}, where loss
and operation errors are detected or even actively corrected at the repeater stations. 

QRs can be classified into three generations depending on the methods
used to overcome loss and operation errors \cite{Gingrich2003,Muralidharan2015a}. 
The first generation QRs \cite{Dur1999,Briegel1998} relies on heralded entanglement generation \cite{Bernien2013} between neighboring repeater stations to overcome loss errors and entanglement purification \cite{Bennett1996,Dur1999} between neighboring and remote repeater stations to correct operation errors. The remote two way classical communication needed between remote stations slows down the key generation rates and requires long lived quantum memories at repeater stations. The second generation QRs relies on heralded entanglement generation between neighboring repeater stations to overcome loss errors and quantum error correction to overcome operation errors \cite{Jiang2008,Munro10,Azuma2015,Pant2016,Epping2016,Li2013}. 
This needs two-way classical communication only between neighboring stations, which can be achieved in parallel. The third generation QR uses only quantum error correction \cite{Grassl1997,Lu2008,Bergmann2016,NC00} to overcome both loss and operation errors \cite{Fowler2010, Zwerger2016, Zwerger2018, Brinew2, Munro2012a,Muralidharan2014,Muralidharan2017,Gingrich2003,Ewert2015,Namiki2016,glaudell2016}.
The third generation QRs are analogous to classical repeaters because
their communication rate depends solely on the time taken to perform
local operations and is a completely one-way protocol, with the potential of reaching ultrafast communication rates \cite{Muralidharan2014,glaudell2016}. Teleportation based error
correction (TEC) \cite{Knill2005} have been introduced at each repeater
station to correct loss and operation errors in the third generation
QRs \cite{Muralidharan2017}. Similar to quantum teleportation, TEC protocol requires encoded Bell state preparation and measurement of logical $X$ and $Z$ operators
of the code. TEC for QR has been generalized to accommodate qudit error correcting
codes such as quantum polynomial codes \cite{Muralidharan2017} using generalized Pauli operators 
that act on a $d$-level system as $X^l|j\rangle = |j+l\rangle$ and $Z^l|j\rangle = \omega^{lj} |j\rangle$, $0\leq i,j\leq d-1$ \cite{Gottesman1999}. The performance of quantum parity codes (QPC) was first studied for one-way QRs \cite{Munro2012a} and the repeater parameters were optimized using a
cost function \cite{Muralidharan2014}. It has been
shown that quantum polynomial codes (QPyCs) \cite{Cleve1999,Muralidharan2017},
which can correct up to a maximum of $50\%$ photon losses \cite{Bennett1997},
can reduce the cost for low operation errors. Moreover, 
continuous variable cat codes can also be used for this generation of QRs \cite{Li2017}. 

So far all proposed third generation QR schemes rely on encoding a single
logical qubit (qudit) into a block of physical qubits (qudits). This
leads to an interesting question of whether the third generation QR can benefit from
error correcting codes encoding multiple logical qubits/qudits, which might enable us to 
further improve the key generation rates. There are efficient ways to construct qubit and qudit quantum error correcting codes from classical linear codes using the Calderbank-Shor-Steane (CSS) construction \cite{NC00}. For third generation QRs, we need CSS codes because the encoded CNOT gate required by the TEC protocol has a transversal implementation \cite{NC00}. For example, a Steane $[[7,1,3]]$ code \cite{NC00} can be constructed from the $[7,4,3]$ classical Hamming code and its dual $[7,3,3]$ code.

Reed-Solomon codes introduced in 1960 \cite{Reed1960} have found spectacular applications \cite{McEliece1981,Wicker1999} for information transmission in the past few decades and has revolutionzed the information technology industry. This motivates us to consider the
construction of quantum Reed-Solomon codes from classical Reed-Solomon codes using the CSS construction and consider their usefulness for quantum communication. More specifically, here, we show that quantum Reed-Solomon codes (QRSC) \cite{Grassl1999,Li2008}
encoding multiple logical qudits can be a promising candidate
for the third generation QRs. Since, we encode multiple logical
qudits into QRSC, one can expect an increase in data transfer rates
compared to encoding schemes where a single logical qubit (qudit)
is being encoded. 

In the following, we first introduce the construction of QRSC \cite{Grassl1999,Li2008}
from classical Reed Solomon codes \cite{Reed1960,McEliece1981,Wicker1999}
using the CSS construction \cite{NC00}. We then show that QRSC of qudit
with dimension $d$ approaches the capacity of the quantum erasure
channel of $d$-level systems. We then describe the application of
QRSC for QRs and study the improvement compared
to QPC and QPyC in terms of the cost coefficient. Finally, we compare the
three generations of QRs assuming QRSC for the third generation and identify
experimental parameter regimes where each generation performs the
best.
\section{Classical Reed-Solomon codes}
A classical Reed Solomon code is defined over a Galois field $GF(d)$,
where $d$ is a prime, which contains a primitive element $\alpha$ such that
$\alpha^{d-1}=1$. $GF(d)=\{0,\alpha,\alpha^{2}...\alpha^{d-2},1\}$.
A Reed Solomon code is defined in the following manner \cite{Reed1960,McEliece1981,Wicker1999}:
Suppose, $c=(c_{0},c_{1},...c_{k-1})$ is a list of information symbols with 
each element taken from $GF(d)$, we define
the polynomial function
\begin{equation}
p_c(x)=c_{0}+c_{1}x+...+c_{k-1}x^{k-1}, 
\end{equation}
which can generate the codewords
\begin{equation}
c=(c_{0},c_{1},...,c_{k-1})\mapsto(p_c(0),p_c(\alpha),...,p_c(\alpha^{d-1})).
\end{equation}
We can obtain $d$ linear equations with $k$ variables as following
\begin{align}
p_c(0) & =c_{0},\\
p_c(\alpha^{j}) & =c_{0}+c_{1}\alpha^{j}+c_{2}\alpha^{2j}+...+c_{k-1}\alpha^{(k-1)j},
\end{align}
where $j\in\left\{ 1,...,d-1\right\} $ and the summation is taken
$mod\,d$. Without loss of generality, we can use the first $k$
equations to solve for the codewords, which gives us the $[d,k,d-k+1]_d$
Reed-Solomon code. We rewrite the above equations in the matrix form
\begin{equation}
(p_c(0),p_c(\alpha),...,p_c(\alpha^{d-1}))=c\cdot G,
\end{equation}
with generator matrix
\begin{equation}
G=\begin{bmatrix}1 & 1 & . & . & . & 1\\
0 & \alpha & \alpha^{2} & . & . & \alpha^{d-1}\\
0^{2} & \alpha^{2} & \alpha^{4} & . & . & \alpha^{2(d-1)}\\
. & . & . & . & . & .\\
. & . & . & . & . & .\\
0^{k-1} & \alpha^{k-1} & \alpha^{2(k-1)} & . & . & \alpha^{(k-1)(d-1)}
\end{bmatrix},
\end{equation}
with the corresponding parity check matrix
as 
\begin{equation}
H=\begin{bmatrix}1 & 1 & . & . & . & 1\\
0 & \alpha & \alpha^{2} & . & . & \alpha^{d-1}\\
0^{2} & \alpha^{2} & \alpha^{4} & . & . & \alpha^{2(d-1)}\\
. & . & . & . & . & .\\
. & . & . & . & . & .\\
0^{d-k-1} & \alpha^{d-k-1} & \alpha^{2(d-k-1)} & . & . & \alpha^{(d-1)(d-k-1)}
\end{bmatrix}.
\end{equation}
 It can be verified that the rows of matrix $H$ are orthogonal
to the rows of $G$. $H$ can also be regarded as the generator matrix of the dual
code $[d,d-k,k+1]_{d}$ \cite{Wicker1999}.

\section{Quantum Reed-Solomon codes}
A QRSC \cite{Grassl1999,Li2008} can be obtained from the CSS construction
of two classical Reed-Solomon codes, namely the $[d,k,d-k+1]_{d}$
code and its dual $[d,d-k,k+1]_{d}$ code, giving us the quantum code $[[d,2k-d,d-k+1]]_{d}$
\cite{Ketkar2006,Guardia2009,Guardia2012} where $2k-d$ logic qudits
of $d$ levels are encoded into $d$ physical qudits. So that as long
as less than (or equal to) $d-k$ qudits are lost, the encoded quantum information
can be retrieved. Therefore, the classical codes 
\begin{eqnarray}
 & C_{1} & =\{p_{c}(0),p_{c}(\alpha),...p_{c}(\alpha^{d-1})|c\in F_{d}^{k}\}\\
 & C_{2} & =\{p_{c}(0),p_{c}(\alpha),...p_{c}(\alpha^{d-1})|c\in F_{d}^{d-k}\}
\end{eqnarray}
fulfil the requirement of CSS construction $C_1 \subset C_2$ and hence they can 
yield the quantum code (i.e. QRSC) 
\begin{equation}
|s_{0},s_{1}...,s_{2k-d-1}\rangle\mapsto\sum_{\textcolor{black}{{c_{d-k+j}=s_{j}},|c\in F_d^k}} |p_{c}(0) p_{c}(1)...p_{c}(\alpha^{d-2})\rangle.
\end{equation}
To understand the above summation, let us consider an example of the $[[3,1,2]]_3$ code with $d=3$,
$k=2$ and $c_{1+j}=s_j$.  \textcolor{black}{The primitive element of $GF(3)$ is 2}. This gives us the codeword
\begin{equation}
|s_0\rangle \mapsto\sum_{\textcolor{black}{c_{1+j}=s_{j}}}|p_{c}(0)p_{c}(1)p_{c}(2)\rangle.
\end{equation} 
For $s_0=0$, we have $c_1=0$, for $s_0=1$, we have $c_1=1$ and for $s_0=2$, we have $c_1=2$. 
The logical states are given by, 
\begin{small}
\begin{eqnarray}
&|0\rangle_L& = \sum_{c_0=0}^2|c_0\rangle|c_0\rangle  |c_0 \rangle = 
\frac{1}{\sqrt{3}} (|000\rangle+ |111\rangle + |222\rangle), \\ 
&|1\rangle_L& = \sum_{c_0=0}^2 |c_0\rangle|c_0+1 \rangle|c_0+2\rangle  = 
\frac{1}{\sqrt{3}} (|012\rangle+ |120\rangle + |201\rangle), \nonumber \\ 
&|2\rangle_L& = \sum_{c_0=0}^2|c_0\rangle |c_0+2 \rangle |c_0+4\rangle= 
\frac{1}{\sqrt{3}} (|021\rangle+ |102\rangle + |210\rangle). \nonumber  
\end{eqnarray} 
\end{small}
Note that the addition is performed modulo 3 here. QPyC is related to the special case of QRSC with $k=(d+1)/2$, for encoding a single logical qudit. Note that QPyC construction does not require the size of the encoding block to be equal to the dimension of qudit $d$ \cite{Cleve1999}, so QPyC is not a subset of QRSC. We will now show how to construct the stabilizers and logical operators of QRSC. 
\subsubsection*{Example 1: $[[3,1,2]]_{3}$ code}
To construct a $[[3,1,2]]_{3}$ code, we use two classical codes, namely, the 
$[3,2,2]_3$ code and the
$[3,1,3]_3$ code. The generator and the parity check matrices for the
$[3,2,2]_3$ code are given by
\[
G=\begin{bmatrix}1 & 1 & 1\\
0 & \alpha & 1
\end{bmatrix},\,H=\begin{bmatrix}1 & 1 & 1\end{bmatrix}
\]
respectively. Similarly the generator and parity check matrices of the $[3,1,3]_3$ code
are given by $H$ and $G$ respectively. The stabilizers of the
the $[[3,1,2]]_{3}$ code are given by, $XXX$ and $ZZZ$, while the logical operators are $X_{L}=IXX^{2}$ and $Z_{L}=IZ^{2}Z$. By multiplying stabilizers 
we obtain equivalent expressions for logical operators
$X_{L}=X^{2}IX,\,XX^{2}I$ and $Z_{L}=ZIZ^{2},\,Z^{2}ZI$. 
If the first qutrit is erased, we can restore the encoded information based on the logical operators 
$X_L = I X X^2$ and $Z_L = I Z^2 Z$, independent of the first qutrit. Similarly, we can restore the encoded information, if the second or third qutrit is erased. Therefore, $[[3,1,2]]_3$ code can correct a single erasure error.

\subsubsection*{Example 2: $[[5,3,2]]_{5}$ code}
To construct a $[[5,3,2]]_{5}$ code, we pick two classical codes, namely $[5,4,2]_5$ code and the $[5,1,5]_5$ code. The generator and parity check matrices of the
$[5,4,2]_5$ code is given by 
\begin{center}
$G=\begin{bmatrix}1 & 1 & 1 & 1 & 1\\
0 & \alpha & \alpha^{2} & \alpha^{3} & 1\\
0 & \alpha^{2} & 1 & \alpha^{2} & 1\\
0 & \alpha^{3} & \alpha^{2} & \alpha & 1
\end{bmatrix}$, $H=\begin{bmatrix}1 & 1 & 1 & 1 & 1\end{bmatrix}$ 
\end{center}
The stabilizer generators of the code are $XXXX$ and $ZZZZ$. The logical operators can also be constructed from the matrices. For example, the $X_L$ operators are given by, $X_{L}^{(1)}=IXX^{2}X^{3}X^{4}$, $X_{L}^{(2)}=IX^{2}X^{4}X^{2}X^{4}$, $X_{L}^{(3)}=IX^{3}X^{2}XX^{4}$.
Other equivalent logical operators can be obtained by multiplying
the stabilizers with these logical operators. The $Z_L$ operators can be obtained in a similar fashion.

\section{Capacity of quantum erasure channel}
For erasure probability $p_{l}$, the capacity of qudit erasure channel
is $1-2p_{l}$ dits/channel use \cite{Bennett1997}. In the following, we show that QRSC can approach this capacity for large $d$, which is associated with both the size of the encoding block and the physical dimension of the d-level system \footnote{We cannot claim
classical Reed-Solomon codes to be capacity achieving on the classical erasure channel because we have a sequence of codes for various values of $d$. However, from our definition of QRSC, there exists a single code for a given value of $d$.}
To justify this claim, we may generally
compute the success probability of error correction for the
$[[d,2k-d,d-k+1]]_{d}$ code with prime $d$ as 
\begin{equation}
\mathrm{P_{success}}=\sum_{j=0}^{d-k}\left(\begin{array}{c}
d\\
j
\end{array}\right)p_{l}^{j}\left(1-p_{l}\right)^{d-j}.
\end{equation}
The failure probability is
\begin{equation}
\mathrm{P_{fail}}=\sum_{j=d-k+1}^{d}\left(\begin{array}{c}
d\\
j
\end{array}\right)p_{l}^{j}\left(1-p_{l}\right)^{d-j},
\end{equation}
which can be rewritten as 
\begin{equation}
\mathrm{P_{fail}}=\sum_{j=0}^{k}\left(\begin{array}{c}
d\\
j
\end{array}\right)p_{l}^{d-j}\left(1-p_{l}\right)^{j},
\end{equation}
Let $k=\left(1-p_{l}\right)d+x$, then $\varepsilon=x/d$ with $x\ll d$.
We have $p_{l}>\frac{d-k}{d}$. According to the Chernoff-Hoeffding theorem
\begin{small}
\begin{eqnarray}
&&\mathrm{P_{success}}|_{1-\frac{k}{d}=p_{l}-\varepsilon}  \leq\left(\left(\frac{p_{l}}{p_{l}-\varepsilon}\right)^{p_{l}-\varepsilon}\left(\frac{1-p_{l}}{1-p_{l}+\varepsilon}\right)^{1-p_{l}+\varepsilon}\right)^{d}
\nonumber \\ 
&&=  e^{-D(p-\epsilon||p)}d
\end{eqnarray}
\end{small}
where
\begin{equation}
D(a||b) = a \mathrm{ln} \frac{a}{b} + (1-a)\mathrm{ln}\frac{1-a}{1-b}
\end{equation}
is the Kullback-Leibler divergence \cite{Kullback1951}. Taking the Taylor expansion of $D$, For $d\rightarrow\infty$ and letting $\varepsilon=x/d\ll1$, we have
\begin{eqnarray}
\mathrm{P_{success}}|_{p_{l}} & \leq & e^{-\varepsilon x/p_{l}}e^{-\varepsilon x/\left(1-p_{l}\right)}
\end{eqnarray}
For $p_{l}<\frac{d-k}{d}$, we have
\begin{eqnarray}
\mathrm{P_{success}}|_{1-\frac{k}{d}=p_{l}+\varepsilon} & = & 1-P_{fail}|_{1-\frac{k}{d}=p_{l}+\varepsilon}.
\end{eqnarray} 
According to the Chernoff-Hoeffding theorem, 
\begin{equation}
\mathrm{P_{fail}}|_{1-\frac{k}{d}=p_{l}+\varepsilon}\leq\left(\left(\frac{p_{l}}{p_{l}+\varepsilon}\right)^{p_{l}+\varepsilon}\left(\frac{1-p_{l}}{1-p_{l}-\varepsilon}\right)^{1-p_{l}-\varepsilon}\right)^{d}.
\end{equation}
For $d\rightarrow\infty$ and let $\varepsilon=x/d\ll1$, we have
\begin{eqnarray}
\mathrm{P_{success}}|_{p_{l}=1-\frac{k}{d}-\varepsilon} & \geq & 1-e^{-(p_{l}+\varepsilon)x/p_{l}}e^{(1-p_{l}-\varepsilon)x/\left(1-p_{l}\right)}. \nonumber \\
\end{eqnarray}
Therefore, for $d\rightarrow\infty$, $P_{success}\approx1$ for $p_{l}-\frac{1-R_{c}}{2}<O(\frac{1}{\sqrt{d}})$,
and $P_{success}=0$ for $p_{l}-\frac{1-R_{c}}{2}>O(\frac{1}{\sqrt{d}})$,
with $R_{c}=\frac{2k-d}{d}$ is the code rate.

\section{One-way quantum repeaters with QRSC}
\begin{figure}[H]
\centering \includegraphics[width=10cm]{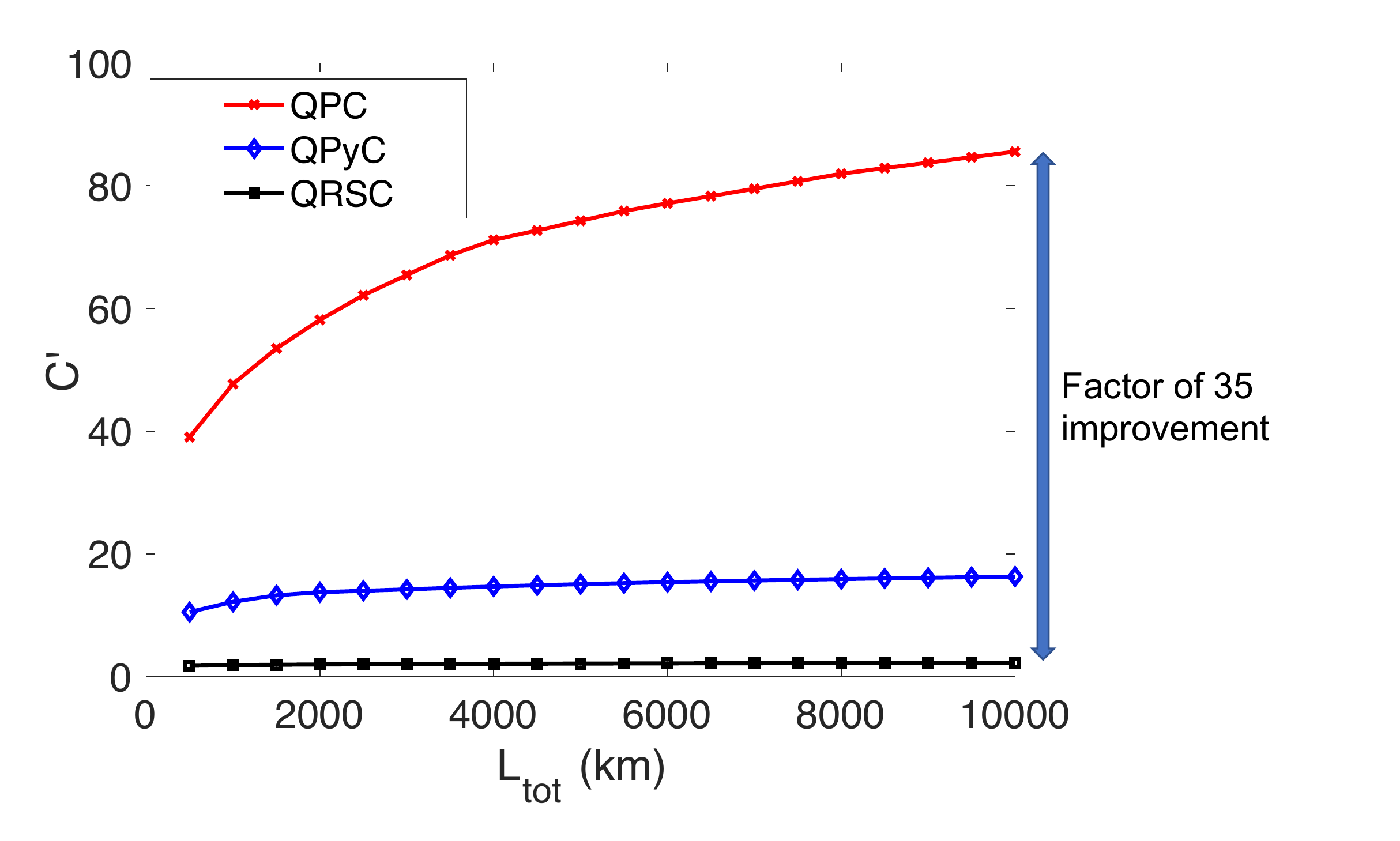} \caption[Comparison between QPC, QPyC and QRSC in the absence of operation
errors]{\textcolor{black}{Comparison between the cost coefficients of quantum Reed Solomon
codes (QRSC) (denoted with black $\square$) with quantum polynomial codes (QPyC) (denoted with  blue $\diamond$) and quantum parity codes (QPC) (denoted with red $\times$) 
 in the absence of operation errors assuming it takes
the same time to create small encoded blocks of qubits (qudits)}.}
\label{fig:compa} 
\end{figure}

One-way QRs use quantum error correction instead of amplification
used in classical repeaters to counter photon loss in propagation.
Here, the quantum state to be transmitted is encoded into an error
correcting code and sent to the neighboring station where a
TEC operation is performed to correct both loss and operation errors.
After the error correction, the signal is retransmitted
to the neighboring station. This is carried out until the encoded
quantum state reaches the receiver. We will first study the error model 
and describe the optimization of QRSC quantum repeaters. There are two kinds of errors that is encountered by QRs which have to be corrected. 
\begin{enumerate}
\item Photon loss errors: The probability that each photon successfully reaches the neighboring repeater station is $\eta_c^2 \times e^{-\frac{L_0}{L_{\text{att}}}}$, where $L_0$ is the repeater spacing, $L_{\mbox{att}} = 20\mbox{km}$ is the attenuation length of the fiber, and $\eta_c$ is the coupling efficiency between fiber and matter qudits. 
\item Operation errors: These include the gate errors, depolarization errors and measurement errors. 
Here, $\epsilon_g$ is the gate error, $\epsilon_d$ is the depolarization error and $\epsilon_m$ is the measurement error. 
The total error measured at the $X$ and $Z$ measurements of the TEC circuit is given by 
$(3\epsilon_g+4\epsilon_d+\epsilon_m)$ \cite{Muralidharan2017}. 
\end{enumerate}
If $x$ photons are lost during the communication, among the rest of the photons that reach the destination, $y$ photons suffer operation errors, as such, the code can correct up to $x + 2y \leq (d-k)$ errors \cite{glaudell2016, Muralidharan2017}. The probability that at least $(d-k)$ photons are received and the encoded state is decoded correctly is given by 
\begin{small}
\begin{eqnarray}
P_{\mathrm{correct(X/Z)}} = \sum_{x=0}^{d-k} \sum_{y=0} ^{\lfloor \frac{d-k}{2} - \frac{x}{2} \rfloor} {d \choose x} { d-x \choose y} \times \nonumber \\ {(1-\eta)}^{x} 
\epsilon_{\mathrm{X/Z}} ^{y} {(1-p_l)}^{d-x} {(1-\epsilon_{\mathrm{X/Z}} )} ^{d-x-y}.
\label{eq:pcorrect}
\end{eqnarray}
\end{small}
The probability that at least $(d-k)$ photons are received and the encoded state is decoded incorrectly is given by 
\begin{small}
\begin{eqnarray}
P_{\mathrm{incorrect(X/Z)}} = \sum_{x=0}^{d-k} \sum_{y=\lceil \frac{d-k}{2} - \frac{x}{2} +0.5 \rceil}
^{(d-k-x)} {d \choose x} { d-x \choose y} \times \nonumber \\ {(1-\eta)}^{x} 
\epsilon_{\mathrm{X/Z}} ^{y} {(1-p_l)}^{d-x} {(1-\epsilon_{\mathrm{X/Z}} )} ^{d-x-y}.
\label{eq:pincorrect}
\end{eqnarray}
\end{small}
By making an assumption that an effective logical error in any one of the QR stations leads to a logical error at the receiver's end, the quantum bit error rate can be defined as \cite{Muralidharan2017}, 
\begin{equation}
Q_{\mathrm{X/Z}} = 1-\frac{{\left[P_{\mathrm{correct(X/Z)}}\right]}^{r}} {{[\text{P}_{\text{success}}]}^r},
\end{equation}
where $r$ is the number of repeater stations. For the two basis protocol for quantum key distribution, the asymptotic secure key generation rate is \cite{Sheridan2010}
\begin{equation}
R = (2k-d)\frac{{[\text{P}_{\text{success}}]}^r} {t_0} \left( \mbox{log}_{2} \mathrm{d} - 2h(Q)\right),
\end{equation}
where $t_0$ is the time taken for quantum gates and measurement and
\begin{eqnarray}
&Q& = \left(\frac{Q_X+Q_Z}{2}\right) \nonumber \\ 
&h(Q)& = -Q\mathrm{log}_2 \frac{Q}{d-1} - (1-Q) \mathrm{log}_2 (1-Q). 
\end{eqnarray}
It is worth mentioning that compared to protocols encoding a single logical qudit \cite{Muralidharan2017}, the boost in the 
key generation rates comes by the factor $(2k-d)$ in logical qudit encoding. 
The two important resources one has to consider for one-way QRs are the number of physical qudits and the time consumption. 
The cost coefficient which is obtained by taking the product of temporal
and physical resources needed for the QR to function provides an excellent
tool to compare QR schemes with different error correcting codes \cite{Muralidharan2014, Muralidharan2017}.
The cost coefficient is the number of qubits required per km to generate
one secure bit per unit time $t_0$. Since we need to compare schemes 
based on qubit encoding (e.g. QPC) and qudit encoding (e.g. QPyC and QRSC), we assume that
each qudit of $d$ levels can be mapped into $\mathrm{log_{2}}d$
qubits. Following Ref. \cite{Muralidharan2017}, we define the cost coefficient for QRSC
\begin{equation}
C'=\frac{2d\mathrm{log}{}_{2}d}{L_{0}R},
\end{equation}
where $2d$ is the number of qudits required for TEC, $L_{0}$ is
the repeater spacing and $R$ is the secure key generation rate
In Fig. \ref{fig:compa}, we compare the performances of QRSC with QPyC and QPC in the absence
of operation errors. It can be seen that one can obtain a factor of
35 reduction in cost for communication up to $\mathrm{L_{tot}}=10,000\mathrm{km}$
by using QRSC instead of QPC.
\section{Comparison of different generations of QRs with QRSC for the third generation}
\begin{figure}[H]
\centering \includegraphics[width=9cm]{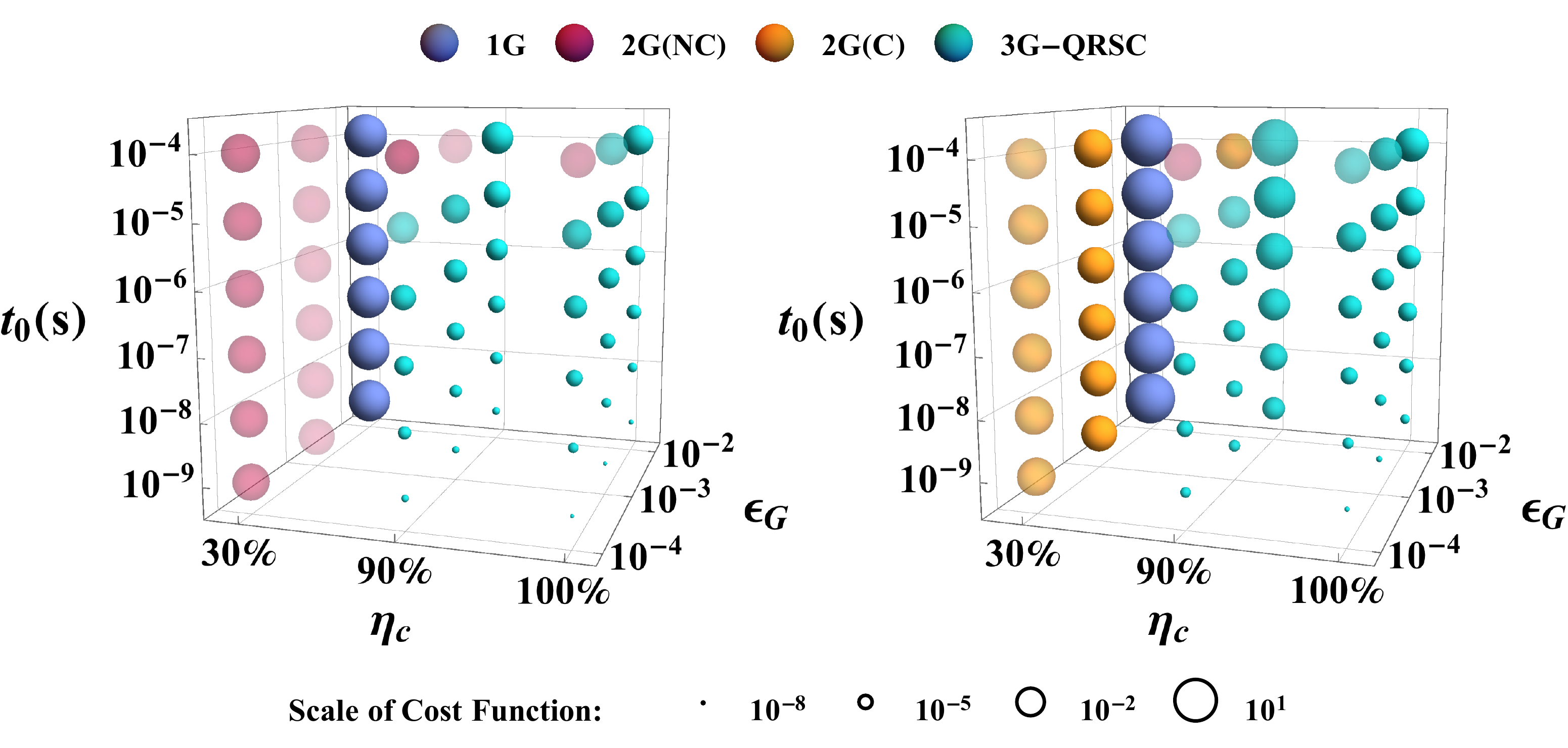}
\caption[Bubble plot comparing various QR generations with QRSC for third-generation.]{The bubble plot comparing various QR protocols in the three-dimensional
parameter space spanned by $\eta_{c}$, $\epsilon_{G}$, and $t_{0}$,
for a) $L_{tot}=1000$km and b) $L_{tot}=10,000$km. The bubble color
indicates the associated optimized QR protocol, and the bubble diameter
is proportional to the log of the cost coefficient.}
\label{fig:bubbleqrsc} 
\end{figure}

The cost coefficient provides an effective tool to compare the three
generations of QRs. For a given set of experimental parameters, we
can identify the quantum repeater generation that yields the minimum
cost and conclude that generation to be the optimum for that set of
experimental parameters. Following Ref. \cite{Muralidharan2015a}, we choose gate error $\epsilon_{G}$, coupling efficiency (between atom and photon) $\eta_{c}$ and operation time
$t_{0}$ for comparison. For first and second generation QRs, we assume that the initial fidelity of Bell pairs generated is $(1-\frac{5}{4}\epsilon_{G})$ with entanglement purification \cite{Muralidharan2015a} and measurement error probability $\epsilon_m = \frac{\epsilon_G}{4}$ obtained through a measurement with an ancilla qubit \cite{Knill05}. 
The three generations of QRs have been
compared using QPC for the third generation with a maximum of 200 qubits
\cite{Muralidharan2015a}. Here, we expand the comparison by including
QRSC for the third generation QRs, with a maximum of $d=23$ which
corresponds to $2d.\mathrm{log}_{2}d\approx$ 200 qubits. This assumes
that we use qubits as elementary building blocks for first and second generations of QRs and QRSC of
qudits for the third generation QRs. 
In Fig.$\,$\ref{fig:bubbleqrsc}, it can be seen that QRSC can correct
a large fraction of erasure errors up to $10^{-2}$. For $\eta_{c}\geq90\%$,
the third generation QRs dominate for all values of $\epsilon_{G}$. For
$\eta_{c}=30\%$, and $\epsilon_{G}=10^{-3},\,10^{-4}$ the third
generation QR cannot correct the loss errors and the second generation
without (with) encoding comes into play. For $\epsilon_{G}=10^{-2}$,
the errors proliferate for first and second generations of QRs and consequently, the
first generation becomes useful here. 

\begin{figure}[h]
\centering \includegraphics[width=8.5cm]{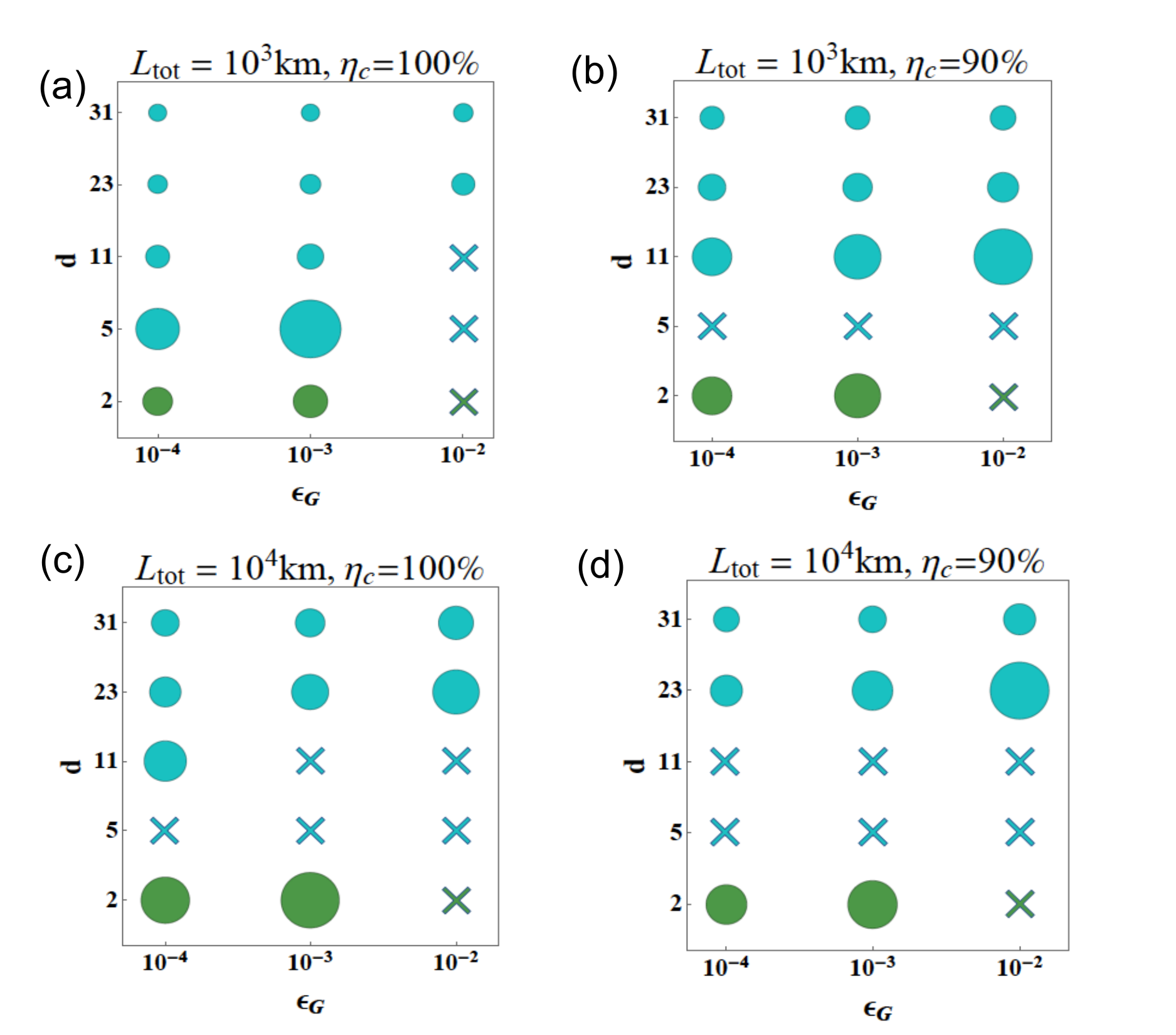} \caption[2D bubble plot showing the $C'$ achievable for various $d$ as a
function of gate error rate$\epsilon_{G}$.]{Bubble plot showing the cost coefficient achievable for various $\epsilon_{G}$
as a function of $d$. $d=2$ corresponds to the case of QPC. $\times$
corresponds to a region not correctable. To beat the cost coefficient
achievable with QPC, one needs to use a QRSC with a large $d$. $d=23$
corresponds to the maximum $d$ searched by the optimization algorithm
for the bubble plot as it is less than 200 qubits. The cost coefficient
does not vary appreciably for $d>31$.}
\label{fig:bubble2dqrsc} 
\end{figure}

Now, we study the variation of cost coefficient with respect to $d$.
In Fig.$\,$\ref{fig:bubble2dqrsc}, the cost function is compared
for $t_{0}=1\,\mathrm{\mu s}$, with $L_{tot}=1000\,\mathrm{km}$
and $10,000\,\mathrm{km}$ respectively. Here, $d\geq3$ corresponds
to QRSC while cases with $d=2$ corresponds to QPC . The results indicate
that for small $\epsilon_G < 10^{-2}$, QRSC can outperform QPC with
a sufficiently large dimension $d$.

\section{Conclusion}
We have investigated quantum Reed-Solomon codes constructed from the CSS construction of
classical Reed-Solomon codes for application of the third generation quantum repeater over long distances. We described the construction of stabilizers
and logical operators for these codes with examples. The rates of
these codes approach channel capacity of the quantum erasure channel for $d$-level systems. For channel dominated
by photon loss errors (with negligible operation errors), QRSC can achieve
a factor of 35 improvement in the cost coefficient compared to QPC. For situations with both loss
and operation errors, we compared the three generations
of QRs including QRSC for the third generation QRs and identified the dimension $d$ where these codes begin to perform better than QPC. As proposed in 
Ref. \cite{Muralidharan2017}, neutral atoms of multilevel systems trapped by photonic nanocrystal cavities \cite{Tiecke14} may provide a potential physical platform to realize qudit based third generation QRs. For future research, it will be interesting to analyze other quantum error correcting codes for
one-way QRs such as quantum Reed-Muller codes \cite{Kumar2016} and quantum
polar codes \cite{Renes2012} and optimize their performance by systematic comparison. 

\section*{Acknowledgments}

We thank Vladimir Malinovsky, Jungsang Kim, Kasper Duivenvoorden, Henry Pfister, 
Hong Tang and Steven Girvin for discussions. This work was supported
by the ARL CDQI, Alfred P. Sloan Foundation (BR-2013-049), ARO
(W911NF-14-1-0011, W911NF-14-1-0563), ARO MURI (W911NF-16-1-0349), AFOSR MURI (FA9550-14-1-0052, FA9550-15-1-0015), NSF (EFMA-1640959) and Packard Foundation (2013-39-273).

\bibliography{collection.bib}

\end{document}